\documentclass[a4paper,11pt]{article}
\pdfoutput=1 

\usepackage{jheppub} 

\usepackage[T1]{fontenc} 

\newcommand{\be}{\begin{eqnarray}}                                             
\newcommand{\ee}{\end{eqnarray}}
\newcommand{\nn}{\nonumber}
\newcommand{\noplus}{}

\newcommand{\tmop}[1]{\ensuremath{\operatorname{#1}}}
\title{\boldmath Non-perturbative to perturbative QCD via the FFBRST}

\author[a,b]{Haresh Raval}
\author[b,1]{Bhabani Prasad Mandal,\note{Corresponding author.}}
\author[a]{Urjit A. Yajnik}

\affiliation[a]{Department of Physics, Indian Institute of Technology, 
	Bombay,\\ Mumbai - 400076, India}
\affiliation[b]{Department of Physics, Institute of Science, Banaras 
Hindu University, \\
Varanasi - 221005, India}

\emailAdd{haresh@phy.iitb.ac.in}
\emailAdd{bhabani.mandal@gmail.com}
\emailAdd{yajnik@phy.iitb.ac.in}

\abstract{Recently a new type of quadratic gauge was introduced in QCD in which the degrees of freedom are suggestive of a phase of abelian dominance. In its simplest form it is also free of Gribov ambiguity. However this gauge is not suitable for usual perturbation theory. The finite field dependent BRST (FFBRST) transformation is a method established  to interrelate generating functionals for different effective versions of gauge fixed field theories. In this paper we propose a FFBRST transformation suitable for transforming the theory in the new quadratic gauge into the standard Lorenz gauge Faddeev-Popov  version of the effective lagrangian. The task is made interesting by the fact the BRST invariance obeyed by the two effective lagrangians are not the same, however suitable extension of the previous procedures accomplishes the required result. We are thus able to identify a field redefinition  to go from a non-perturbative phase of QCD to perturbative QCD.
}

\begin{document} 
\maketitle
\flushbottom

\section{Introduction}
Extensions of the usual Lorenz gauge by including the next order terms quadratic in gauge fields have been studied in several contexts
\cite{8,9,10,11,12,13}. In \cite{Raval:2014pxa} it was shown that a purely quadratic gauge condition without the linear terms leads to a suggestive 
effective lagrangian giving masses to off-diogonal gluons. The consequences of such a condition to lifting the Gribov ambiguity were further studied in 
\cite{Raval:2014pxa} and \cite{Raval:2016sxe}. 
The new type of quadratic gauge condition is at first introduced as follows,
\begin{align} \label{eq:0}
H^a [ A^{\mu} ( x) ] =
A^a_{\mu} ( x) A^{\mu a} ( x) = f^a ( x) ; \  \text{  for each $a$ }
\end{align}
where $f^a(x)$ is an arbitrary function of $x$.  Several proposals to establish abelian dominance in the infrared (IR) use what is called Abelian 
Projection \cite{ap}. 
Such and other algebraic gauges are usually non-covariant. But as introduced in ~\cite{Raval:2014pxa} the above gauge is in fact covarint . 
The gauge prima facie is  an ambiguity free gauge as it is algebraic in nature. Thus, the quadratic gauge shares the same  
property of being free of  Gribov copies as the axial gauges $n_\mu A^{\mu a} = f^a(x)$ and the flow 
gauge $\alpha A_0^a = \nabla. \vec{A^a}$~\cite{Chan:1985kf} despite being non linear. The 
Faddeev-Popov determinant in this gauge is given by
\be 
\det\left(\frac{\delta (A^{a \epsilon}_\mu A^{\mu a \epsilon})}{\delta \epsilon^b}\right)= \det\left(2A^a_\mu( \partial^\mu \delta^{ab}-g f^{acb}A^{\mu c})\right),
\ee
Therefore,
the resulting effective Lagrangian density contains gauge fixing and ghost terms as follows,
\begin{align} \label{eq1}
\mathcal{L}_{\tmop{GF}}+ \mathcal{L}_{\tmop{ghost}} =&- \frac{1}{2 
	\zeta}\displaystyle\sum\limits_{a}  ( A^a_{\mu} A^{\mu a})^2  - 
2\displaystyle\sum\limits_{a}\overline{c^a}A^{\mu a} ( D_{\mu} c)^a , 
\end{align}
where $ \zeta$ is an arbitrary gauge fixing parameter and $(D_{\mu} c)^a = 
\partial_\mu c^a - g f^{a b c} A_\mu^b c^c$. 
Now onwards, we shall drop the summation symbol,
but the  summation over an index $a$ will be 
understood  when it  appears repeatedly, including when repeated \textit{thrice} as 
in the ghost terms above.
In particular,
\begin{equation}\label{ghost}
- \overline{c^a} A^{\mu a} ( D_{\mu} c)^a 
= - \overline{c^a} A^{\mu a} \partial_{\mu} c^a + g f^{a b c} 
\overline{c^a} c^c A^{\mu a} A^b_{\mu}
\end{equation}
where the summation over indices $a$, $b$ and $c$ each runs independently over $1$ to $N^2-1$. We should note that ghost Lagrangian does not have kinetic
terms and hence the ghosts do not propagate in this theory and make no loop contributions. They act like auxiliary fields, but playing an important role in the IR.
With this understanding, we write the full effective Lagrangian density in this quadratic gauge as
\begin{eqnarray} \label{eq:Leff}
\mathcal{L}_{Q} &=&- \frac{1}{4} F^a_{\mu \nu} F^{\mu \nu a}
\noplus - \frac{1}{2 \zeta}  ( A^a_{\mu} A^{\mu a})^2 - 2\overline{c^a}
A^{\mu a} ( D_{\mu} c)^a  \nn \\
&=& - \frac{1}{4} F^a_{\mu \nu} F^{\mu \nu a}
\noplus + \frac{\zeta}{2 }  F^{a2}+ F^a A^a_{\mu} A^{\mu a} -2 \overline{c^a}
A^{\mu a} ( D_{\mu} c)^a  ,
\end{eqnarray}
where the field strength $F^a_{\mu \nu}= \partial_{\mu}A^a_{\nu}(x)- \partial_{\nu}A^a_{\mu}(x)-g 
f^{abc} A^b_{\mu}(x)A^c_{\nu}(x)$ and in the second version the $F^a$ are a set of auxiliary fields called Nakanishi-Lautrup fields\cite{Weinberg:1996kr}. 
As shown in ~\cite{Raval:2016sxe}, the Lagrangian is BRST  invariant\cite{Becchi:1974md, Becchi:1975nq}
which is essential for the ghost independence of the green functions and unitarity of 
the $S$-matrix. These issues were studied in ref.~\cite{Raval:2014pxa}

The form of the second term of the expression  \eqref{ghost} appearing in the ghost lagrangian contains ghost bilinears multiplying terms quadratic in
gauge fields. Hence if the non-propagating ghosts are assumed to be frozen they amount to a non-zero mass matrix for the gluons.
To strengthen this connection it is necessary to assume that the vacuum corresponds to ghost condensation. This was achieved  through 
introducing a Lorenz gauge fixing term for one of the diagonal gluons, in addition to the purely quadratic terms of Eq. \eqref{eq:0}.  
This gauge fixing gives the propagator to the corresponding ghost field. Using this ghost propagator, one can give nontrivial vacuum values 
to bilinears $ \overline{c^a} c^c $ within the framework Coleman-Weinberg mechanism as described 
in \cite{Raval:2014pxa}. We shall revisit the point in the next section also.

The resulting mass matrix for the gluons has $N(N - 1)$ non-zero eigenvalues only and thus has 
nullity $ N -1$. Thus, the $N(N - 1)$ off-diagonal gluons acquire masses and the rest $N-1$ diagonal gluons remain massless.
The massive off-diagonal gluons are presumed to provide evidence of Abelian dominance, which is a  signature of quark confinement. 
This and other phenomena that emerge in this gauge, such as the avoidance of Gribov ambiguity were studied explicitly 
in ~\cite{Raval:2014pxa} ~\cite{Raval:2016sxe}.
Quark confinement  and Gribov ambiguity are important 
non-perturbative issues. And this gauge therefore proves to be important in studying  non-perturbative regime of QCD.

The finite field dependent BRST (FFBRST) transformation was introduced for first time in Ref. 
\cite{13} by integrating infinitesimal usual BRST transformations. Such FFBRST transformations 
have exactly the same form as the infinitesimal ones, with the difference that the infinitesimal global 
anti-commuting parameter is replaced by an anti-commuting but finite 
parameter dependent on space time fields, but with no explicit dependence on space time coordinates. 
The meaning of "finite anti-commuting parameter" is that if we calculate the Green's functions for 
such parameters
between vacuum and a state with gauge and ghost fields we get finite values as opposed to 
infinitesimal values.
Being finite in nature FFBRST  transformation does not leave the path integral measure invariant 
even though other properties of usual infinitesimal BRST transformation are intact. Thus the 
generating functional to a BRST invariant theory is not invariant under FFBRST. Jacobian of such 
finite transformation provides a non-trivial factor which depends on FFBRST parameter. 

Due to this 
non-trivial Jacobian FFBRST transformations are simultaneously field redefinitions as well
as BRST transformations on the fields being redefined. They are thus capable 
of connecting generating functionals of two different BRST invariant theories    and have been used 
to study different gauge field theoretic models with various
effective actions \cite{sdj1,rb,susk,deg,ffanti, bss,fs,sudhak,ffanti1}.
In this paper we construct an appropriate FFBRST transformation to establish the connection at the level of 
generating functionals between the recently introduced quadratic gauge  with substantial 
implications  in the non-perturbative 
QCD \cite{Raval:2014pxa,Raval:2016sxe} and the familiar Lorenz gauge which is  suitable to describe the 
perturbative QCD.

\section{Connecting two different regimes}

As discussed in the introduction,  the main non perturbative result of the quadratic gauge was established with the help of additional gauge fixing for one of the gluons.  
The presence of this additional gauge fixing does reintroduce the Gribov ambiguity for this component but this is the price to be paid for an explicit demonstration of effective masses for the off diagonal gluons.
Hence,  our aim here is to connect the generating functional  corresponding to the effective actions in 
quadratic gauge with additional Lorenz gauge fixing for one of the gluons
to  that in the usual Lorenz gauge through the technique of the FFBRST transformation. To 
do so, we  write  the effective action of the quadratic gauge with additional gauge fixing for one of the gluons which is as follows
\begin{eqnarray} \label{eqnew}
\mathcal{S}_{eff}&=& \mathcal{S}_{Q} + \int d^4x\left[\frac{\xi}{2 }  (G^{3})^2+G^3\partial^\mu A_\mu^3 -\overline{d^3}	\partial^{\mu } ( D_{\mu} d)^3\right] \\
&=&  \mathcal{S}_{Q}+\int d^4x\left[ \frac{\xi}{2 }  (G^{3})^2+G^3\partial^\mu A_\mu^3 -\overline{d^3}	\Box d^3\right] \ \ \nn
\end{eqnarray}
where $\Box$ stands for $\partial^\mu \partial_\mu$ and a set of additional fields $G^3,d^3, \overline{d^3}$ correspond to the additional Lorenz gauge of the diagonal gluon $A^3$ and  the ghosts $\overline{ d^3}, d^3$ are treated as $SU(3)$ singlets. 

As a first approach it is easy to see that the action in Eq.~\eqref{eqnew} is invariant under the following nilpotent  BRST transformation 
\begin{eqnarray}\label{tra}
\begin{split}
\delta c^d =& \ \frac{\delta\omega}{2}f^{dbc}c^bc^c\\
\delta \overline{c^d } = &\frac{\delta \omega}{g}F^d \\
\delta A^d_\mu=&\ \frac{\delta \omega}{g}(D_\mu c)^d \\
\delta F^d =& \ 0\\
\delta G^3=&\ 0\\
\delta  \overline{d^3 }=&\ 0\\
\overline{d^3 }  \delta \Box d^3=& \ \frac{\delta \omega}{g}G^3 \partial^\mu D_\mu c^3
\end{split}
\end{eqnarray}
where $\delta \omega $ infinitesimal, anticommuting and global parameter. This set of transformations differs from the usual BRST transformation in the composite form of the last of the Eq.s~\eqref{tra}. However these transformations are not useful for FFBRST technique since the transformations of $G^3$ and $\overline{d^3}$ are trivial. Therefore, we introduce a new set of BRST transformations under which the action~\eqref{eqnew} is also invariant, as follows 
\begin{eqnarray}\label{tra1}
\begin{split}
\delta c^d =& \ \frac{\delta\omega}{2}f^{dbc}c^bc^c\\
\delta \overline{c^d } = &\frac{\delta \omega}{g}F^d \\
\delta A^d_\mu=&\ \frac{\delta \omega}{g}(D_\mu c)^d \\
\delta F^d =& \ 0\\
\delta G^3=&\ \frac{\delta\omega}{g} \Box d\\
\delta  \overline{d^3 }=&\ \frac{\delta\omega}{g}(\partial_\mu A^{\mu 3}+ \xi G^3)\\
\overline{d^3 }  \delta \Box d^3=& \ \frac{\delta \omega}{g}G^3 \partial^\mu D_\mu c^3
\end{split}
\end{eqnarray}
We see that  now this set of transformations has become applicable for FFBRST technique as $\delta G^3, \delta  \overline{d^3 }$ are non zero. 
Further, the last three transformations in Eq.~\eqref{tra1} are not nilpotent, but they satisfy following higher degree closed algebra
\be\label{nl0}
\overline{d^3}\delta^2(\underline{\overline{d^3}\delta \Box d^3} )&=&0\nn \\
\overline{d^3}\delta^2(\overline{d^3} \ \underline{\delta^3 \overline{d^3}} )&=&0.
\ee
The remaining one can be easily derived from one of these algebras.
Thus, we are prompted to restore the nilpotency, to the extent possible.
To do so, we express the effective action~\eqref{eqnew} in terms of a new auxiliary field $B^3$, 
\be\label{af}
\mathcal{S}_{eff} &=&  \mathcal{S}_{Q}+\int d^4x\left[ \frac{\xi}{2 }  (G^{3})^2+G^3\partial^\mu A_\mu^3 -\overline{d^3}	\Box d^3\right]  \nn\\
&=& \mathcal{S}_{Q}+\int d^4x\left[ \frac{-1}{2 \xi}  (B^{3})^2+ B^3G^3+G^3\partial^\mu A_\mu^3 -\overline{d^3}	\Box d^3\right]
\ee
The action~\eqref{af} is invariant under following transformations
\begin{eqnarray}\label{tra2}
\begin{split}
\delta c^d =& \ \frac{\delta\omega}{2}f^{dbc}c^bc^c\\
\delta \overline{c^d } = &\frac{\delta \omega}{g}F^d \\
\delta A^d_\mu=&\ \frac{\delta \omega}{g}(D_\mu c)^d \\
\delta F^d =& \ 0\\
\delta G^3=&\ \frac{\delta\omega}{g} \Box d\\
\delta B^3=& \ -\frac{\delta\omega}{g}\partial^\mu D_\mu c^3\\
\delta  \overline{d^3 }=&\ \frac{\delta\omega}{g}(\partial_\mu A^{\mu 3}+  B^3)\\
\overline{d^3 }  \delta \Box d^3=& \ \frac{\delta \omega}{g\xi}B^3 \partial^\mu D_\mu c^3
\end{split}
\end{eqnarray}
The last three transformation rules satisfy the following algebra
\be\label{nl}
\delta^2 B^3 &=& 0 \nn\\
\delta^2 \overline{d^3} &=&0\\
\delta(\underline{\overline{d^3}\delta \Box d^3})&=&0\nn
\ee
We see that first two rules of Eq.~\eqref{nl}  are nilpotent and the last one is `almost' nilpotent now. It is interesting to compare Eqs.~\eqref{nl0} and \eqref{nl}.
Thus, we see that the introduction of $B^3 $ has made a substantial  difference in the algebra of the transformations,
with the novel feature of the algebra of the transformation having been made nilpotent through the introduction of an auxiliary field.
We shall next achieve the stated connection using these unusual transformations~\eqref{tra2} in the FFBRST technique. 

Now we briefly outline the procedure for the passage from the  BRST transformations to the FFBRST transformations. We start  with making the 
infinitesimal global parameter $\delta\omega$ field dependent  by introducing a numerical parameter $\kappa \ (0\leq \kappa\leq 1) $ and 
making all the fields  $\kappa$ dependent
such that  $\phi (x,\kappa=0)=\phi(x) $ and $\phi (x,\kappa=1)=\phi^\prime(x) $, the transformed field.
The symbol $\phi$ generically describes all the fields $A,c,\overline{c},F,d^3,\overline{d}^3,B^3,G^3$.  The BRST transformation in Eq. \eqref{tra2} is then written as
\be\label{infb}
d\phi = \delta_b[\phi(x,\kappa)]\Theta^\prime(\phi(x,\kappa))\ d\kappa
\ee
where $\Theta'$ is a finite field dependent anti-commuting parameter and 
$\delta_b[\phi(x,\kappa)]$ is the form of the transformation for the corresponding field as in 
Eq.~\eqref{tra2}. The
FFBRST is then constructed by integrating Eq. \eqref{infb} from $\kappa=0$ to $\kappa=1$ as \cite{13}
\be\label{ffbrst}
\phi^\prime\equiv \phi(x,\kappa=1)=\phi (x,\kappa=0)+\delta_b[\phi(0)]\Theta [\phi(x)]
\ee
where $\Theta[\phi(x)] =\int_0^1 d\kappa^\prime\Theta^\prime[\phi(x,\kappa)] $. Like usual BRST transformation,  FFBRST
transformation leaves the effective action in Eq.~\eqref{af} invariant. However, since the transformation parameter is field dependent in nature, 
FFBRST transformation does not leave the path integral measure, ${\cal D}\phi $ invariant and produces a non-trivial Jacobian factor $J$. 
This $J$  can further be cast as a local  functional of fields, $ e^{iS_J}$ (where the $S_J$ is the action representing the Jacobian factor $J$) 
if the following condition is met \cite{13}
\be \label{con}
\int { \cal D }\phi (x,\kappa) \left [\frac{1}{J}\frac{dJ}{d\kappa}-i\frac{dS_J}{d\kappa}\right ]e^{i(S_J+\mathcal{S}_{eff})}=0.
\ee

Thus the procedure for FFBRST may be summarised as (i)  calculate the infinitesimal change in 
Jacobian, $\frac{1}{J}
\frac{dJ}{d\kappa} d\kappa $ using 
\begin{equation}
\frac{J(\kappa)}{J(\kappa+d\kappa) }= 1-\frac{1}{J(\kappa)}\frac{dJ(\kappa)}{d\kappa}d\kappa
= \sum_\phi \pm \frac{\delta\phi(x,\kappa+d\kappa)}{\delta\phi(x,\kappa)}
\end{equation}
for infinitesimal BRST transformation,$+$ or $-$ sign is for Bosonic or Fermion nature of the field  $\phi$ respectively 
(ii)  make an ansatz for $S_J$, (iii)  then  prove the Eq. (\ref{con})
for this ansatz and finally (iv) replace $J(\kappa)$ by $e^{iS_J}$ in the generating functional
\begin{equation}
W=\int {\cal D}\phi (x) e^{iS_{eff}(\phi)} = \int {\cal D}\phi (x,\kappa) J(\kappa)e^{iS_{eff}(\phi (x,\kappa))} .
\label{ww}
\end{equation}
Setting $\kappa=1$, this would then provide the new effective action $ S^\prime_{eff}=S_J+S_{eff}$.

Now we proceed to construct a FFBRST transformation with an appropriate parameter to connect 
the generating functionals in the quadratic gauge with additional Lorenz gauge for  the diagonal gluon $A^3$ and the Lorenz gauge.  We construct the finite 
field dependent parameter as 
\be\label{th}
\Theta'[\phi(k)]=-i \int d^4 x \left [ \overline{c^a}\Big( \gamma_2A^a_{\mu} A^{\mu a}+ \gamma_3 \partial_{\mu} A^{\mu a}\Big)+\gamma_1 \overline{d^3}G^3\right]
\ee
The
$ \gamma_1, \gamma_2$ and $\gamma_3$ are constant parameters and $\Theta'^2=0$. Group index $a$ is summed over. This FFBRST transformation is particularly different among others~\cite{sdj1,rb,susk,deg,ffanti,bss,fs}
due to the unique form of transformations~\eqref{tra2} and by the fact that the field dependent parameter in Eq.~\eqref{th} contains two  
ghosts $\overline{c},\overline{d}$ with two different transformation properties unlike others where there is only one ghost. 
We now calculate the change in the Jacobian $\frac{1}{J}\frac{dJ}{d\kappa} $ due to the FFBRST with the parameter in Eq.~\eqref{th}, under which the measure changes  $\mathcal{D}\phi(\kappa) \rightarrow J(\kappa)\mathcal{D}\phi(\kappa)$ as
\be
\frac{1}{J}\frac{dJ}{dk} &=& - \frac{1}{g}\int d^4x\Big((D_\mu c)^a \frac{\delta \Theta'}{\delta A_\mu^a}+ \partial^\mu(D_\mu c)^a \frac{\delta \Theta'}{\delta \partial^\mu A_\mu^a} - \frac{\delta (\Theta'f^{abc}c^bc^c)}{\delta c^a}
-F^a \frac{\delta \Theta'}{\delta \overline{c^a}}\nn\\&-&(\partial_\mu A^{\mu 3}+ B^3)\frac{\delta \Theta'}{\delta \overline{d^3}}+ \Box d^3 \frac{\delta \Theta'}{\delta G^3}\Big)\nn\\
&=&\frac{i}{g}\int d^4x\Big(2\gamma_2\overline{c^a}(D_\mu c)^a A^{\mu a} + \gamma_3\overline{c^a}\partial^\mu(D_\mu c)^a- F^a\big(\gamma_2 A^a_{\mu} A^{\mu a}+\gamma_3  \partial_{\mu} A^{\mu a}\big)\nn\\&-& \gamma_1 G^3 (\partial_\mu A^{\mu 3}+ B^3)+\gamma_1\overline{d^3}\Box d^3\Big)
\ee
Since $\frac{1}{J}\frac{dJ}{dk}$ does not contain terms with $\Theta'$ as multiplicative factor, the $\kappa$ dependence in $S_J(\kappa)$ is multiplicative~\cite{13}. This implies that the fields in the ansatz for the $S_J$ can be taken to be $\kappa$ independent. With this fact in mind, 
we  make the following ansatz for the $S_J$ to compensate the Jacobian contribution of FFBRST transformation 
\be\label{sj}
S_J[\phi,\kappa]&=& \frac{1}{g}\int d^4 x\Big(   \alpha_1 (\kappa)F^a A^a_{\mu} A^{\mu a}  + 2\alpha_2 (\kappa)\overline{c^a}A^{\mu a} ( D_{\mu} c)^a  +  \alpha_3 (\kappa) F^a \partial_{\mu} A^{\mu a} \nn\\& +&\alpha_4 (\kappa) \overline{c^a}\partial^{\mu } (D_{\mu} c)^a  + \alpha_5(\kappa)G^3(\partial^\mu A_\mu^3+B^3)+ \alpha_6(\kappa)\overline{d^3}\Box d^3  \Big)
\ee
where $\alpha_j(\kappa), j=1,\cdots 6$, are arbitrary functions with initial condition $\alpha_i(\kappa=0)=0$ while the fields themselves are $\kappa$ independent.
We calculate,
\be
i\frac{dS_J}{dk}&=& \frac{i}{g}\int d^4x \Big[ \dot{\alpha_1}F^a A^a_{\mu} A^{\mu a}+2 \dot{\alpha_2}\overline{c^a}A^{\mu a} ( D_{\mu} c)^a+\dot{\alpha_3}F^a \partial_{\mu} A^{\mu a}+ \dot{\alpha_4}\overline{c^a}\partial^{\mu} (D_{\mu} c)^a\nn\\&+&\dot{\alpha_5}G^3(\partial^\mu A_\mu^3+B^3)+\dot{\alpha_6}\overline{d^3}	\Box d^3 
\ee
In order to satisfy the condition in Eq.~\eqref{con},  the following equation must be obeyed
\be
&&\int {\cal D}\phi[x,\kappa]\Big [F^a A^a_{\mu} A^{\mu a}(-\gamma_2-\dot{\alpha_1})+2\overline{c^a}A^{\mu a} ( D_{\mu} c)^a(\gamma_2-\dot{\alpha_2})\
+F^a \partial_{\mu} A^{\mu a}(-\gamma_3-\dot{\alpha_3})\nn\\&+&\overline{c^a}\partial^{\mu} (D_{\mu} c)^a(\gamma_3-\dot{\alpha_4})
+ (-\gamma_1-\dot{\alpha_5})G^3(\partial^\mu A_\mu^3+B^3)+(\gamma_1-\dot{\alpha_6})\overline{d^3}\Box d^3
\Big ] e^{i(S_{eff}+S_J)}=0,\nn\\
\ee
which gives the following relation among parameters
\begin{eqnarray}\label{tr}
\begin{split}
\dot{\alpha_1}&= -\dot{\alpha_2}= -\gamma_2\\
\dot{\alpha_3}&=-\dot{\alpha_4}= -\gamma_3\\
\dot{\alpha_5}&=-\dot{\alpha_6}=-\gamma_1.\\
\end{split}
\end{eqnarray}
The Eqs.~\eqref{tr}  have the obvious solutions 
\begin{eqnarray}\label{para}
&&\alpha_1=-\alpha_2  = -\gamma_2\kappa; \ \alpha_3 = -\alpha_4= -\gamma_3\kappa, \nn\\&&\alpha_5=-\alpha_6=-\gamma_1 \kappa
\end{eqnarray}
We choose 	the arbitrary parameters   $\gamma_1 =1, \gamma_2 =1$, 
$\gamma_3=-1$ in Eq. \eqref{para}.
Thus, the additional Jacobian contribution at $\kappa=1$ is
\be\label{s}
S_J&=&\int d^4 x\Big(-F^a A^a_{\mu} A^{\mu a} 
+ 2\overline{c^a}A^{\mu a} ( D_{\mu} c)^a 
+  F^a \partial_{\mu} A^{\mu a} - \overline{c^a}\partial^{\mu } (D_{\mu} c)^a\nn\\&-& G^3(\partial^\mu A_\mu^3+B^3) + \overline{d^3}\Box d^3 \Big).
\ee
Adding this Jacobian contribution, $S_J$ to the $\mathcal{S}_{eff}$ in Eq.~\eqref{af} we obtain at $\kappa=1$ the Lorenz gauge as follows
\be
\mathcal{S}_{eff}+S_J 
&=&  \int d^4 x\left[ -\frac{1}{2\xi }  (B^{3})^2+\frac{\zeta}{2 }  F^{a2}+   F^a \partial_{\mu} A^{\mu a} - \overline{c^a}\partial^{\mu} ( D_{\mu} c)^a\right]\nn\\
&=& S_{L}
\ee	
Here the term $\frac{1}{\xi } (B^{3})^2$ is redundant which can be put to zero by using EOM for $B^3$.
Now, we may further apply second FFBRST such that $\zeta \rightarrow \zeta'$ in the same Lorenz gauge by well known methods \cite{13}.
We may summarize this symbolically as the conversion from one 
theory to another,
\be
Z_{eff} =\int {\cal D}\phi e^{i\mathcal{S}_{eff} } \stackrel{FFBRST}{\longrightarrow} \int {\cal D}\phi ^\prime (\kappa)e^{i(\mathcal{S}_{eff}
	+S_J)} \  = \int {\cal D}\phi ^\prime e^{iS_L }= Z_L,
\ee  Thus, we have connected two theories with two different regimes of 
applicability. This is a connection also between theories with and without propagating ghosts.

\section{Conclusion}
The spirit of BRST invariance was to establish the unitarity of the S-matrix in gauge theories whose gauge fixed versions contain ghost degrees of 
freedom. This technique was substantially extended in the FFBRST approach 
to permit field redefinitions transforming the effective action with one possible gauge fixing to that of another. In some of the recent earlier work the interesting features
of a purely quadratic gauge condition without the usual Lorenz condition have been studied and shown to lead to several interesting properties of the
non-perturbative QCD vacuum in the IR limit. At first site the effective degrees of freedom entering here, the off diagonal gluons with masses, appear 
unrelated to those entering the perturbation theory calculations and which are compatible with the elegant UV properties of Yang-Mills theories.  In this paper
we have resorted to the FFBRST technique to establish a direct formal connection between the two varieties of the QCD effective lagrangians. Several technical 
difficulties are encountered in this process and it has required us to make suitable extensions to the FFBRST method. In particular a new auxiliary field is required 
to ensure nilpotency of the modified BRST transformations. The resulting field redefinitions which connect the degrees of freedom capturing the IR behaviour of
QCD vacuum with those of  the UV version suitable to perturbative computations need to be studied further. 
Also, the extensions of the FFBRST technique proposed here can be put to use for other similar problems.

\acknowledgments

This work is partially supported by Department of Science and Technology, Govt. of India under National Postdoctoral Fellowship scheme
 with File no. `PDF/2017/000066'. 
One of us (BPM) acknowledges the support from Physics Department, IIT-Bombay for a visitation during which the work was initiated.


\end{document}